\documentclass{article}

\usepackage[margin=1.25in]{geometry}
\usepackage{wrapfig}
\usepackage{times}
\usepackage{soul}
\usepackage{url}
\usepackage[utf8]{inputenc}
\usepackage[small]{caption}
\usepackage{graphicx}
\usepackage{floatrow}
\usepackage{amsmath}
\usepackage{booktabs}
\usepackage{algorithm}
\usepackage{algorithmic}
\usepackage{microtype}
\usepackage{subfigure}
\usepackage{amssymb}
\usepackage{amsthm}
\usepackage{cases}
\usepackage{amsfonts}       
\usepackage{nicefrac}       
\usepackage{sidecap}
\usepackage{tikz}
\usepackage{enumerate}
\usepackage{subfiles}
\usepackage{authblk}
\usepackage{natbib}

\newtheorem{lemma}{Lemma}

\newtheorem{problem}{Problem}
\newtheorem{theorem}{Theorem}
\newtheorem{proposition}{Proposition}

\usetikzlibrary{decorations.pathreplacing}

\renewcommand{\cite}[1]{\citep{#1}}

\title{An Efficient Evolutionary Algorithm for Minimum Cost Submodular Cover}
\author[1,*]{Victoria G. Crawford}
\affil[1]{University of Florida}
\affil[*]{vcrawford01@ufl.edu}
\date{}
\begin{document}

\maketitle

\begin{abstract}
  In this paper, the Minimum Cost Submodular Cover problem is studied, which
  is to minimize a modular cost function such that the monotone submodular benefit function is
  above a threshold. For this problem,
  an evolutionary algorithm EASC is introduced that achieves a constant, bicriteria
  approximation in expected polynomial time; this is the first polynomial-time
  evolutionary approximation algorithm for Minimum Cost Submodular Cover.
  To achieve this running time, ideas motivated by submodularity and monotonicity
  are incorporated into the evolutionary process, which likely will extend to
  other submodular optimization problems.
  In a practical application, EASC is demonstrated to outperform the greedy
  algorithm and converge faster than competing evolutionary algorithms for this
  problem.
\end{abstract}

\section{Introduction}

\label{section:introduction}
  A function $f:2^S\to\mathbb{R}_{\geq 0}$ defined on subsets of a ground set $S$
  is monotone submodular if it possesses the following two properties:
  \begin{itemize}
    \item[i.] For all $A\subseteq B\subseteq S$, $f(A)\leq f(B)$ (monotonicity).
    \item[ii.] For all $A\subseteq B\subseteq S$
    and $x\notin B$, $f(A\cup\{x\})-f(A) \geq f(B\cup\{x\})-f(B)$ (submodularity).
  \end{itemize}
  Monotone submodular set functions and optimization
  problems are found in many applications in machine learning and data mining.
  In this paper, the 
  NP-hard Minimum Cost Submodular Cover Problem (MCSC) is considered, which is defined as follows.
  \begin{problem}[Minimum Cost Submodular Cover (MCSC)]
    Let $S$ be a ground set of size $n$. Let
    $c:2^S\to\mathbb{R}_{\geq 0}$ be a modular\footnote{The function $c$ is modular if $c(X)=\sum_{x\in X}c(\{x\})$ for all $X\subseteq S$.}
    function such that $c(X)=0$ if and only if $X=\emptyset$,
    and $f:2^S\to\mathbb{R}_{\geq 0}$ be monotone submodular.
    Given a threshold $\tau \leq f(S)$, MCSC is to find
    $\text{argmin}\{c(X): X\subseteq S, f(X) \geq\tau\}.$
    The function $c$ is called the cost, while $f$ is called the benefit.
  \end{problem}%
  Applications of MCSC
  include data summarization \cite{mirzasoleiman2015,mirzasoleiman2016},
  active set selection \cite{norouzi2016}, recommendation systems \cite{guillory2011},
  and viral marketing in social networks \cite{kuhnle2017}.

  The standard greedy
  algorithm\footnote{The greedy algorithm is discussed in Section \ref{appendix:greedy} of the Appendix.}
  is an effective, efficient approximation algorithm for MCSC \cite{wolsey1982analysis}; however, once the
  greedy solution has been obtained, it is unclear how it could be improved if more
  computational resources are available.
  Therefore, it is of interest
  to employ methods that can improve the solution quality
  at the expense of more runtime, while
  maintaining a worst-case
  guarantee.
  For this reason, an evolutionary algorithm has recently been proposed for
  MCSC \cite{qian2015constrained}.

  %
  Although random search methods
  such as evolutionary algorithms (EA) can find better quality
  solutions in practice,
  it is difficult to analyze the approximation quality of evolutionary algorithms.
  The algorithm of \citeauthor{qian2015constrained} is able to improve upon the greedy solution in practice,
  but requires expected exponential time\footnote{Time is measured in number of
  evaluations of $f$ and $c$, as is commonly done \cite{badanidiyuru2014fast}.}
  to have a worst-case guarantee similar to that of the greedy algorithm.
  Furthermore, no evolutionary algorithm exists
  in prior literature for MCSC that achieves such an approximation ratio in polynomial time.

  \subsection{Contributions}
  This paper presents the novel algorithm, EASC (Evolutionary Algorithm for Submodular Cover, Alg. \ref{alg:bins}),
  which is the first polynomial-time evolutionary
  algorithm for MCSC with constant, bicriteria approximation
  ratio:
  EASC finds a
  solution $A$ such that $f(A)\geq (1-\epsilon)\tau$ and
  $c(A)\leq(\ln(1/\epsilon)+1)c(A^*)$,
  where $A^*$ is an optimum solution and $\epsilon\in(0,1)$ is an input parameter.
  The expected time is $\mathcal{O}(n^3((c_{max}/c_{min})\ln(1/\epsilon))^2),$ where
  $c_{max}$ and $c_{min}$ are the maximum and minimum cost of a single element $s \in S$,
  respectively.
  If $c_{max}/c_{min}$ is bounded by a polynomial in $n$
  and $\epsilon$ is a constant, then EASC finds a near-feasible solution to MCSC
  with a constant approximation ratio in expected polynomial time.

  In contrast to existing EAs that have been analyzed for submodular optimization problems
  \cite{qian2015subset,qian2015constrained,qian2017subset},
  EASC is not a generic EA for multi-objective optimization.
  Instead, EASC takes advantage of the structure of monotone
  submodular functions to quickly
  strengthen its population. A key idea in EASC is that the range $[0,\tau]$
  is discretized into \textit{bins} and
  subsets of $X\subseteq S$ are mapped to bins based on the value of $f(X)$.
  Solutions within a bin compete with one another using a novel measure of
  cost-effectiveness. Both the bin structure and notion of cost-effectiveness
  are designed to take advantage of monotonicity and submodularity.
  It is likely that these ideas have potential to be applied to monotone submodular
  optimization problems other than MCSC.
%
%
%

  EASC is experimentally evaluated on instances of the Influence Threshold Problem (IT)
  \cite{goyal2013,kuhnle2017} on real social network
  datasets. EASC is compared to both the greedy algorithm as well as the existing EA
  that has been analyzed for MCSC, POM \cite{qian2015constrained}. Both EASC and
  POM are able to find better solutions than the greedy algorithm on the problem
  instances, which demonstrates
  the value of EAs for MCSC. In addition, EASC is shown to converge faster than
  POM on some instances.

  \paragraph{Organization.}
  Related work is first discussed in Section \ref{section:relatedwork}.
  Then, EASC is described in detail in Section \ref{section:algorithm}. Theoretical results on the approximation
  ratio of EASC are presented in Section \ref{section:approximation}.
  Finally, the application and an experimental analysis of EASC is given
  in Section \ref{section:experiments}.

  \paragraph{Notation.}
  The following notation will be used throughout the paper.
  For $x\in S$, define $c(x)=c(\{x\})$ and $f(x)=f(\{x\})$.
  Define $c_{min} = \min_{x\in S}c(x)$, and $c_{max} = \max_{x\in S}c(x)$.
  Let $f_{\tau}(X) = \min\{f(X), \tau\}$ for all $X\subseteq S$.
  The notation for marginal gain is shortened to
  $\Delta f(X,x) = f(X\cup\{x\})-f(X)$
  for $X\subseteq S$ and $x\in S$.
  Finally, $\exp(a)$ denotes the exponential function $a \mapsto e^a$.

  %


\subsection{Related Work}
  \label{section:relatedwork}
  Evolutionary algorithms (EAs) have previously been analyzed for submodular optimization
  problems \cite{friedrich2015maximizing,qian2015constrained,qian2015subset,qian2017subset,friedrich2018heavy}.
  In general, these EAs work by maintaining a population of non-dominating\footnote{
  A solution $X$ is dominated by a solution $Y$ if $c(Y)\leq c(X)$ and $f(Y)\geq f(X)$.
  The domination is strict if at least one of the inequalities is strict.
  } solutions. Iteratively, a random solution from the population is selected and mutated.
  If the new solution is not strictly dominated by
  an existing solution in the population, it is kept\footnote{In addition, the EA may require that
  the new solution meet some requirement such as the cost being beneath a bound.
  } in the population and solutions dominated by the new solution are removed from the population.
  These EAs are quite generic and apply broadly to
  multi-objective optimization problems. In contrast, EASC is designed specifically for MCSC.

  \citeauthor{friedrich2014maximizing} (\citeyear{friedrich2014maximizing}) and
  \citeauthor{qian2015subset} (\citeyear{qian2015subset}) analyzed similar
  EAs for the problem of maximizing a monotone, submodular function with respect to
  a cardinality constraint $k$.
  \citeauthor{friedrich2014maximizing} obtained as good an approximation
  ratio as the greedy algorithm in expected $\mathcal{O}(n^2(\log(n)+k))$ time and
  \citeauthor{qian2015subset} in expected $\mathcal{O}(k^2n)$ time for this problem.


  \citeauthor{qian2015constrained} (\citeyear{qian2015constrained}) analyzed an EA
  for MCSC called POM (Pareto Optimization Method) \cite{qian2015constrained}.
  \citeauthor{qian2015constrained}
  proved that the population of POM would contain an
  $H_{c\tau}$\footnote{$c$ is the minimum real number making
  $cf(X)$ for all $X\subseteq S$ and $c\tau$ integers, and the $c\tau$ harmonic number is
  $H_{c\tau} = \sum_{j=1}^{c\tau}1/j$.} $= O( \log (c\tau) )$  approximate solution
  for MCSC in
  $\mathcal{O}(Nn(\log(n)+\log(c_{max})+N)))$
  expected time,
  where $N$ is the number of distinct $f$ values in $[0,\tau)$.
  In order that the approximation guarantees of POM be in expected polynomial time,
  the number of distinct values of
  $f$ in the region of $[0,\tau)$ must be bounded by a polynomial.
  However, this is not a realistic assumption for many applications in machine learning and data mining,
  where $f$ is real-valued and easily takes on exponentially many values in the region $[0,\tau)$
  \cite{kuhnle2017,mirzasoleiman2015}.

  The $N$ in the number of expected time comes from the population size
  of POM.
  Hence in POM, the population can get quite large, which in turn affects the expected time before
  the approximation ratio is reached.
  A similar issue arises when an EA for the dual problem of
  MCSC
  is analyzed \cite{qian2017subset}.
  EASC does not have this problem as its population size
  is always $\mathcal{O}((c_{max}/c_{min})\ln(1/\epsilon)n)$.
  There exist results on approximating a set of non-dominating solutions with a set of
  smaller size \cite{laumanns2002combining,horoba2009additive}.
  In fact, \citeauthor{laumanns2002combining} and
  \citeauthor{horoba2009additive} both describe approaches of binning solutions
  that serves a similar purpose to the bins in EASC, though the bins in EASC are
  quite different; among other reasons,
  solutions that dominate others in the population of EASC are possible.
  However, it is not clear
  that the approaches described by \citeauthor{laumanns2002combining} and
  \citeauthor{horoba2009additive}
  could be done efficiently in this context nor could result in
  approximation ratios in expected polynomial time.

\begin{algorithm}[tb]
  \caption{Evolutionary Algorithm for MCSC (EASC)}
  \label{alg:bins}
  \begin{algorithmic}
    \STATE {\bfseries Input:} MCSC instance parameters $f:2^S\to\mathbb{R}_{\geq 0}$, $c:2^S\to\mathbb{R}_{\geq 0}$,
    and $\tau$, \texttt{bin} parameters $\delta\in[1-c_{min}/c(A^*),1-c_{min}/c(S)]$ and $\epsilon\in(0,1)$,
    and number of iterations $T \in \mathbb{Z}_{\geq 0}$.

    $t=1, \mathcal{B}=\{\emptyset\}$

    \texttt{bin} $\leftarrow$ The bin function induced by $\tau, \delta, \epsilon$

    $\prec \leftarrow$ The comparison operator induced by $\tau, \texttt{bin}$

    \WHILE {$t \leq T$}
    \STATE {
      $X$ uniformly randomly chosen from $\mathcal{B}$

      $X'$ = mutate($X$)

      \IF {$\exists Y \in\mathcal{B}$ such that \texttt{bin}$(Y) = $ \texttt{bin}$(X')$}
      \IF {$Y \prec X'$}
      \STATE $\mathcal{B} = \mathcal B \setminus \{ Y \} \cup \{ X' \}$
      \ENDIF
      \ELSE
      \STATE $\mathcal{B} = \mathcal B \cup \{ X' \}$
      \ENDIF

      $t = t + 1$
    }
    \ENDWHILE

    \end{algorithmic}
  \end{algorithm}


\pagebreak
\section{Evolutionary Algorithm for MCSC (EASC)}
\label{section:algorithm}
  In this section, the algorithm EASC (Evolutionary Algorithm for Submodular Cover) is introduced.
  Pseudocode for EASC can be found in Algorithm \ref{alg:bins}.
  EASC is designed for finding good approximate solutions to
  instances of MCSC efficiently.
  As will be shown in Section \ref{section:approximation},
  if the input parameter $\epsilon$ is constant and $c_{max}/c_{min}$ is bounded by
  a polynomial, then
  EASC provides a near-feasible solution to MCSC with a constant
  approximation ratio in expected polynomial time.

  Fundamental to EASC is a mapping from $2^S$ to $\mathcal{O}((c_{max}/c_{min})\ln(1/\epsilon)n)$ bins;
  \textit{bin} $j$ is associated with the subinterval
  $$\left[ \left(1 - \delta^j \right) \tau , \left(1 - \delta^{j+1} \right) \tau \right) \subseteq [0,\tau],$$ and a subset
  $X\subseteq S$ is mapped into the bin where $\min\{f(X),\tau\}$ falls.
  The population $\mathcal{B} \subseteq 2^S$, which is a set of subsets of $S$,
  contains at most one subset of $S$ per bin.
  The bins are discussed in more detail in Section \ref{section:bin}.

  The input parameters $\epsilon\in(0,1)$ and $\delta\in[1-c_{min}/c(A^*),1-c_{min}/c(S)]$, where
  $A^*$ is an optimal solution to the instance,
  determine the number of bins and the intervals for each bin.
  Lower $\epsilon$ and $\delta$ values result in less bins,
  and hence a smaller population size.
  To find a $\delta$ in the required range, $\delta$ can be set to $1-c_{min}/B$ where $B$
  is an upper bound on $c(A^*)$ such that $B\leq c(S)$. In the experiments in Section \ref{section:experiments},
  $B$ is set to the cost of the greedy solution.

  The number of iterations of EASC is determined by the input parameter $T$.
  At each iteration, EASC chooses a solution $X \in \mathcal B$
  to mutate to $X'$, under the
  mutation process described in Section \ref{section:mutation}.
  If it is not the case that $X'$ is mapped to a bin
  with a better solution according to comparison $\prec$, $X'$ is
  added to $\mathcal B$, and the weaker solution is removed, if any.
  The comparison operator $\prec$ is discussed in Section \ref{section:compare}.

  The structure of the bins and the comparison operator $\prec$ are motivated by
  monotonicity and submodularity, as described in
  Sections \ref{section:bin} and \ref{section:compare}.

\subsection{The Bin Function}
\label{section:bin}
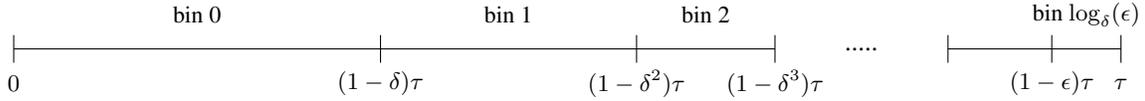
\begin{figure*}
\centering
\begin{tikzpicture}[scale=0.92]
\draw (0,0) -- (11,0);

\node at (0,-0.5) {\small$0$};
\draw (0,-0.2) -- (0,0.2);

\node at (2.65,0.5) {\small bin 0};

\node at (5.3,-0.5) {\small$(1-\delta)\tau$};
\draw (5.3,-0.2) -- (5.3,0.2);

\node at (7.15,0.5) {\small bin 1};

\node at (9,-0.5) {\small$(1-\delta^{2})\tau$};
\draw (9,-0.2) -- (9,0.2);

\node at (10,0.5) {\small bin 2};

\node at (11,-0.5) {\small$(1-\delta^{3})\tau$};
\draw (11,-0.2) -- (11,0.2);

\node at (12.25,0) {.....};

\draw (13.5,0) -- (16,0);

\draw (13.5,-0.2) -- (13.5,0.2);


\node at (15,-0.5) {\small$(1-\epsilon)\tau$};
\draw (15,-0.2) -- (15,0.2);

\node at (15.5,0.5) {\small bin $\log_{\delta}(\epsilon)$};

\node at (16,-0.5) {\small$\tau$};
\draw (16,-0.2) -- (16,0.2);
\end{tikzpicture}
\caption{The region $[0,\tau]$ is discretized into $\log_{\delta}(\epsilon)+1$ bins in EASC. Solutions
$X\subseteq S$ are mapped into the bin corresponding to the region where $\min\{f(X),\tau\}$ falls.
$\mathcal{B}$ contains at most 1 subset of $S$ for each bin.}
\label{fig:bins}
\end{figure*}

In EASC, $[0,\tau]$ is discretized into $\log_{\delta}(\epsilon)+1$ intervals associated
with bins.
Every $X\subseteq S$
is mapped to the bin where $\min\{f(X), \tau\}$ falls:
The function \texttt{bin} takes $X\subseteq S$ and returns a bin number in
$\{0,...,\log_{\delta}(\epsilon)\}$ as follows:
\begin{align*}
  \texttt{bin}(X) =
    \begin{cases}
        i &\text{if } (1-\delta^i)\tau \leq f(X) < (1-\delta^{i+1})\tau \\
          &\text{for an } i \in \{0,...,\log_{\delta}(\epsilon)-1\} \\
        \log_{\delta}(\epsilon) &\text{if } f(X) \geq (1-\epsilon)\tau.
     \end{cases}
\end{align*}
The bins are depicted in Figure \ref{fig:bins}.
The population $\mathcal{B}$ in EASC contains at
most one solution for each bin, and therefore is bounded in size by $\log_{\delta}(\epsilon)+1$.
A solution $X\in\mathcal{B}$ that maps to bin $\log_{\delta}(\epsilon)$ (the final bin) is near-feasible:
$f(X)\geq(1-\epsilon)\tau$. It is the solution mapped to this bin that will give the approximation ratio
in expected polynomial time.

Using the fact that $\delta\leq 1-c_{min}/c(S)$, the total number of bins is bounded as follows.
\begin{proposition}
  \label{proposition:bins}
  The number of bins is at most
  \begin{align*}
    \frac{c_{max}}{c_{min}}\ln\left(\frac{1}{\epsilon}\right)n + 1.
  \end{align*}
\end{proposition}
\begin{proof}
  It is the case that
  \begin{align*}
    \log_{\delta}(\epsilon) &= \frac{\ln(\epsilon)}{\ln{\delta}}
    = \frac{\ln(1/\epsilon)}{-\ln{\delta}}
    \leq \frac{\ln(1/\epsilon)}{1-\delta}
    \leq \frac{c(S)}{c_{min}}\ln\left(\frac{1}{\epsilon}\right).
  \end{align*}
  Since $c(S)\leq c_{max}n$, the result follows.
\end{proof}
The motivation behind the interval assignment of each bin
comes from the greedy algorithm
for MCSC. Suppose the sequence of elements $a_1,...,a_k$ is chosen by
the greedy algorithm for the instance of MCSC. Let $A_i=\{a_1,...,a_i\}$. It is the case\footnote{See
Proposition \ref{proposition:greedy} of Section \ref{appendix:greedy} in the Appendix.}
that for $i<k$ the marginal gain at each step
is lower bounded as follows:
\begin{align*}
  f(A_{i+1}) - f(A_i) \geq \frac{c(a_i)}{c(A^*)}\left(\tau - f(A_i)\right)
\end{align*}
where $A^*$ is an optimal solution to the instance of MCSC. If $\delta\geq 1-c_{min}/c(A^*)$
the region of each bin mimics this marginal gain. Intuitively, the bins can be thought of like steps in
the greedy algorithm. EASC holds on to the best solution for each step.

\subsection{Comparison Operator $\prec$}
\label{section:compare}
If two solutions in $\mathcal{B}$ map to the same bin, then the \textit{weaker} solution is
removed. Weaker is determined by the comparison operator $\prec$.
$\prec$ uses a novel measure of cost-effectiveness, $\phi$, in order to compare solutions.
Let $X\subseteq S$. If $\texttt{bin}(X)=0$ or $\texttt{bin}(X)=\log_{\delta}(\epsilon)$, then $\phi(X)=c(X)$.
Otherwise
\begin{align*}
  \phi(X) = c(X)/\ln\left(\frac{\tau}{\tau - f(X)}\right).
\end{align*}
Notice that lower $\phi$ means better cost-effectiveness.
Then $Y \prec X$ if and only if $\texttt{bin}(X)=\texttt{bin}(Y)$ and $\phi(X) < \phi(Y)$.


\begin{figure}
  \label{fig:compare}
  \centering
\includegraphics[width=0.45\textwidth]{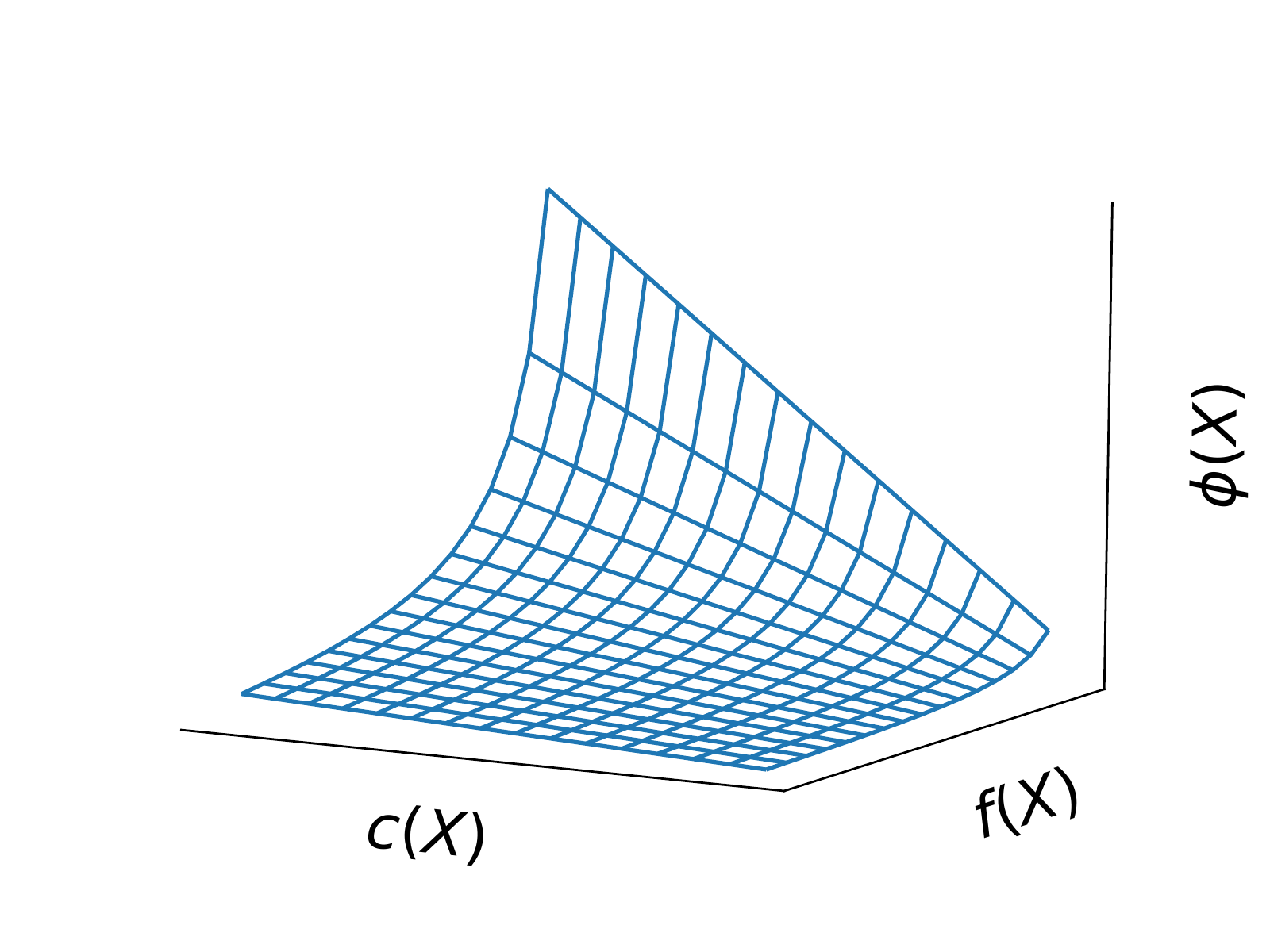}
\caption{An illustration of the cost-effectiveness $\phi$ for $X$ where $\texttt{bin}(X)\in\{1,...,\log_{\delta}(\epsilon)-1\}$.
On both the $c(X)$ and $f(X)$ axis, left is greater. Lower values of $\phi(X)$ means \textit{more} cost-effective.}
\end{figure}

Figure 2 illustrates the cost-effectiveness $\phi(X)$ for varying values of
$c(X)$ and $f(X)$ for $\texttt{bin}(X)\in\{1,...,\log_{\delta}(\epsilon)-1\}$.
Lower values of $c(X)$ and higher values of $f(X)$ result in lower $\phi(X)$ (which means
\textit{more} cost-effective). But as $f(X)$ decreases, differences in $c(X)$ are amplified
in $\phi(X)$. Therefore, $c$ matters more when comparing solutions in lower bins, and in
higher bins (excluding the last) $f$ matters more. This encourages solutions in $\mathcal{B}$
to rise up (via mutation) to the final bin, where a $(1- \epsilon)$-feasible solution is held.

Like the bins discussed in Section \ref{section:bin}, the motivation for $\prec$
is the greedy algorithm for MCSC. Again, consider the sequence of elements chosen by the greedy
algorithm
$a_1,...,a_k$ and
let $A_i=\{a_1,...,a_i\}$. For $i < k$ it is the case\footnote{See Proposition
\ref{proposition:greedy} of Section \ref{appendix:greedy} in the Appendix.} that
\begin{align*}
  c(A_i)/\ln\left(\frac{\tau}{\tau - f(A_i)}\right) \leq c(A^*).
\end{align*}
Therefore at every iteration the greedy algorithm has a solution $A_i$ where
$\phi(A_i)\leq c(A^*)$. In a sense, a solution $X\subseteq S$ such that $\phi(X)\leq c(A^*)$
has as good of cost-effectiveness as solutions picked by the greedy algorithm. $\prec$
ensures that if $\mathcal{B}$ contains a solution $X$ such that $\phi(X)\leq c(A^*)$, then
$X$ cannot be replaced with a solution that is less good in that sense.

\subsection{Mutation of Elements in $\mathcal{B}$}
\label{section:mutation}
Random mutation of solutions in the population occur in the same fashion as existing Pareto
optimization algorithms \cite{qian2015subset}.
At each iteration of EASC, an element $X\in\mathcal{B}$ is chosen uniformly
randomly to be mutated. $X$ is mutated into $X'$ as follows:
Every $x\in X$ is removed from $X$ with independent probability $1/n$.
Every $x\notin X$ is added to $X$ with independent probability $1/n$. The number of elements expected to
change from $X$ to $X'$ is 1.

$X'$ is added to $\mathcal{B}$ if there does not exist a $Y\in\mathcal{B}$ mapping to the same
bin as $X'$ such that
$X' \prec Y$. If no such $Y$ exists, then $X'$ is added to
$\mathcal{B}$ and any existing solution in the bin of $X'$ is removed.

%


\section{Approximation Results}
  \label{section:approximation}

  In this section, the number of iterations before EASC contains a near-feasible
  solution for MCSC with an approximation ratio of $\ln(1/\epsilon)+1$ is analyzed,
  where $\epsilon\in(0,1)$ is an input parameter of EASC. Each iteration of EASC involves exactly one evaluation each of $f$ and $c$.
  If time is measured in
  evaluations of $f$ and $c$, as is commonly done \cite{badanidiyuru2014fast},
  then the expected time is a constant times the expected number of iterations.

  The approximation guarantee in
  Theorem \ref{theorem:main} is a \textit{bicriteria approximation guarantee},
  which means that both the feasibility
  constraint\footnote{The set $X$ is feasible iff $f(X) \ge \tau$.} and the minimum cost are approximated. Algorithms with bicriteria
  approximation guarantees have previously been considered for submodular optimization problems \cite{iyer2013submodular}.

  If $\epsilon$ is assumed to be a constant and $c_{max}/c_{min}$ bounded
  by a polynomial in $n$, then Theorem \ref{theorem:main} shows that EASC
  finds a near-feasible
  solution with a constant approximation ratio in expected polynomial iterations.

    \begin{theorem}
      \label{theorem:main}
      Suppose that we have an instance of MCSC with optimal solution $A^*\neq\emptyset$,
      and EASC is run indefinitely with input $\epsilon\in(0,1)$ and
      $\delta\in [1-c_{min}/c(A^*), 1-c_{min}/c(S)]$.
      Then $\mathcal{B}$ contains a set $A$ in bin $\log_{\delta}(\epsilon)$
      such that $f(A) \geq (1-\epsilon)\tau$ and
      \begin{align*}
        c(A) \leq \left(\ln\left(\frac{1}{\epsilon}\right)+1\right)c(A^*),
      \end{align*}
      where $A^*$ is an optimum solution,
      in expected number of iterations at most
      \begin{align*}
        en\left(\left(\frac{c_{max}}{c_{min}}\right)\ln\left(\frac{1}{\epsilon}\right)n+1\right)^2.
      \end{align*}
    \end{theorem}

  Once a solution that fits the criteria of Theorem \ref{theorem:main} appears in bin
  $\log_{\delta}(\epsilon)$, it cannot be replaced by one that does not since the
  comparison operator $\prec$ compares based on only $c$ in the last bin.
  Notice that Theorem \ref{theorem:main} does not contradict the optimality of
  the $\ln(n)$-approximation ratio for the set cover problem \cite{feige1998},
  since the guarantee is bicriteria.
  The same bicriteria approximation guarantee in Theorem \ref{theorem:main} holds for
  the greedy algorithm\footnote{This result was originally proven for an influence
  application by \citeauthor{goyal2013} (\citeyear{goyal2013}) but holds for general MCSC.
  See Proposition \ref{proposition:greedy} of Section \ref{appendix:greedy} in the Appendix.}.

  The proof of Theorem \ref{theorem:main} tracks $\textit{cost-effective solutions}$
  in $\mathcal{B}$ over the duration of EASC.
  A set $X$ is cost-effective if it satisfies one of the following (mutually exclusive)
  conditions:
  \begin{enumerate}[i.]
    \item $\texttt{bin}(X)<\log_{\delta}(\epsilon)$ and $\phi(X)\leq c(A^*)$.
    \item $\texttt{bin}(X)=\log_{\delta}(\epsilon)$ and $c(X) \leq \left(\ln(1/\epsilon)+1\right)c(A^*)$.
  \end{enumerate}
  Once a cost-effective set is in the final bin, a solution that meets the criteria
  of Theorem \ref{theorem:main} is in the population.
  By design, EASC never replaces a cost-effective solution in its population with
  one that is not cost-effective. In addition, because of the requirement that
  $\delta\geq 1-c_{min}/c(A^*)$, the bins are structured tightly enough
  so that there is a significant probability that cost-effective solutions mutate
  into cost-effective solutions in strictly greater bins.
  Together, these points enable EASC to contain a cost-effective solution in its
  final bin in polynomial expected iterations.
%

  The following lemmas will be used to prove Theorem \ref{theorem:main}.
  The lemmas are not novel to this work, but have previously been used to analyze
  the approximation guarantee of the greedy algorithm for MCSC \cite{goyal2013}. Proofs of
  the lemmas are included in Section \ref{appendix:lemmas} of the Appendix.

  \begin{lemma}
    \label{lemma:gain}
    Suppose that we have an instance of MCSC with optimal solution $A^*\neq\emptyset$.
    Let $X\subseteq S$ and $x^* = \text{argmax}_{x\in S}\Delta f_{\tau}(X,x)/c(x)$.
    Then
    \begin{align*}
      \tau - f_{\tau}(X\cup\{x^*\}) \leq \left(1-\frac{c(x^*)}{c(A^*)}\right)(\tau-f_{\tau}(X)).
    \end{align*}
  \end{lemma}

  \begin{lemma}
    \label{lemma:lastelt}
    Suppose that we have an instance of MCSC with optimal solution $A^*\neq\emptyset$. Let
    $X\subseteq S$ such that $f(X) < \tau$
    and $x^* = \text{argmax}_{x\in S}\Delta f_{\tau}(X,x)/c(x)$.
    Then $c(x^*)\leq c(A^*)$.
  \end{lemma}

\begin{proof}[Proof of Theorem \ref{theorem:main}]
\setcounter{equation}{0}
Recall that notation is defined in Section \ref{section:introduction}.
For brevity, let $r=\log_{\delta}(\epsilon)$ be the final bin.

There always exists at least one cost-effective solution in $\mathcal{B}$:
The empty set is cost-effective since $\phi(\emptyset)=c(\emptyset)=0 < c(A^*)$, and
the empty set is never removed from $\mathcal{B}$ because
there does not exist $Y\subseteq S$ such that $c(Y) < 0 = c(\emptyset)$.
Note that this means if a solution is removed from $\mathcal{B}$,
it can be assumed that it did not correspond to bin 0.

Define an infinite sequence $\ell_t$, $t\in\{1,2,...\}$, where
$\ell_t$ is the max value in $\{0,...,r\}$ such that there exists a cost-effective
solution $X\in\mathcal{B}$ where \texttt{bin}$(X)=\ell_t$ at the beginning of iteration
$t$ of EASC.

\paragraph{Part One.}
First, it is shown that the sequence $\ell_t$ is non-decreasing.
Let $X$ be the cost-effective
set corresponding to $\ell_t$ at the beginning of iteration $t$.
If $X$ is not removed from $\mathcal{B}$
during the $t$th iteration, then clearly $\ell_t \leq \ell_{t+1}$.

Suppose $X$ is removed from $\mathcal{B}$
during the $t$th iteration. Then $X$ was replaced with
$X'$ such that $X \prec X'$ and $\texttt{bin}(X)=\texttt{bin}(X')$.
Let $b=\texttt{bin}(X)=\texttt{bin}(X')$. As explained above, $b\neq 0$.
Suppose $b=r$. Then $X \prec X'$ implies that
$c(X') < c(X) \leq \left(\ln(1/\epsilon)+1\right)c(A^*)$. Therefore $X'$ is cost-effective.
If $b\in\{1,...,r-1\}$. Then $X\prec X'$ implies that
$\phi(X') < \phi(X) \leq c(A^*)$ and hence $X'$ is also cost-effective.
In both of these cases, $\ell_t = \ell_{t+1}$.

\paragraph{Part Two.}
Second, it is shown that if $\texttt{bin}(X)<r$ and
\begin{align*}
  x^* = \text{argmax}_{x\in S}\frac{\Delta f_{\tau}(X,x)}{c(x)},
\end{align*}
then $\texttt{bin}(X\cup\{x^*\}) > \texttt{bin}(X)$.

Let $a=\texttt{bin}(X)$.
Lemma \ref{lemma:gain} and that $\delta\geq 1-c_{min}/c(A^*)$ implies
\begin{align*}
  f_{\tau}(X\cup\{x^*\}) \geq (1-\delta)\tau + \delta f_{\tau}(X).
\end{align*}
By definition of the bins $f_{\tau}(X)\geq (1-\delta^{a})\tau$, and therefore
\begin{align*}
  f_{\tau}(X\cup\{x^*\}) \geq (1-\delta^{a+1})\tau.
\end{align*}
Since $a < r$, it is the case that $\texttt{bin}(X\cup\{x^*\}) \geq a+1$.

\paragraph{Part Three.}
Third, it is shown that if $X$ is cost-effective, $\texttt{bin}(X) < r$, and $x^*$ defined as in Part Two,
then $X\cup\{x^*\}$ is cost-effective. Let $a=\texttt{bin}(X)$ and $b=\texttt{bin}(X\cup\{x^*\})$.
To show the cost-effectiveness of $X\cup\{x^*\}$, four cases are analyzed based on the values of $a$ and $b$.

Case (i): $b<r$ and $a=0$. In this case, $X=\emptyset$ as explained at the beginning of the proof, and
$X\cup\{x^*\}=\{x^*\}$.
Lemma \ref{lemma:gain} states that
\begin{align*}
  \tau - f_{\tau}(x^*) \leq \left(1-\frac{c(x^*)}{c(A^*)}\right)\tau
  \leq \exp\left(-\frac{c(x^*)}{c(A^*)}\right)\tau
\end{align*}
which can be re-arranged to see that $\phi(x^*)\leq c(A^*)$ and hence $X\cup\{x^*\}$ is cost-effective.
%

Case (ii): $b < r$ and $a > 0$. Since $b, a < r$ it is the case that
$f_{\tau}(X\cup\{x^*\}) = f(X\cup\{x^*\})$ and $f_{\tau}(X) = f(X)$.
Lemma \ref{lemma:gain} gives that
\begin{align*}
  \tau - f(X\cup\{x^*\}) \leq \exp\left(-\frac{c(x^*)}{c(A^*)}\right)(\tau-f(X)).
\end{align*}
Using the upper bound on $\tau - f(X)$ given by re-arranging $\phi(X)\leq c(A^*)$
implies that
\begin{align*}
  \tau - f(X\cup\{x^*\}) \leq \exp\left(-\frac{c(X\cup \{x^*\})}{c(A^*)}\right)\tau
\end{align*}
which may be re-arranged to see that $X\cup\{x^*\}$ is cost-effective.

Case (iii): $b=r$ and $a > 0$. $X$ being cost-effective and $a < r$ imply that
$c(X\cup\{x^*\}) =$
\begin{align*}
  c(X) + c(x^*)
  \leq \ln\left(\frac{\tau}{\tau-f(X)}\right)c(A^*) + c(x^*).
\end{align*}
By Lemma \ref{lemma:lastelt}, $c(x^*)\leq c(A^*)$.
Therefore $X\cup\{x^*\}$ is cost-effective.

Case (iv): $b=r$ and $a=0$. Then $X=\emptyset$ as explained at the beginning of the
proof.
Then $c(X\cup\{x^*\})=c(x^*)\leq c(A^*)$ by Lemma \ref{lemma:lastelt}, and therefore
$X\cup\{x^*\}$ is cost-effective.

\paragraph{Part Four.}
It is now shown that if at iteration $t$, the cost-effective set $X$ associated with
$\ell_t < r$ is mutated into $X\cup\{x^*\}$, then $\ell_t < \ell_{t+1}$.

Suppose $X$ is mutated into $X\cup\{x^*\}$ on iteration $t$.
$X\cup\{x^*\}$ is cost-effective by Part Three.
Let $b=\texttt{bin}(X\cup\{x^*\})$.
By Part Two, $b > \ell_t$.
If there does not exist $Y\in\mathcal{B}$ at the beginning of iteration $t$ such that
$\texttt{bin}(Y)=b$, then $X\cup\{x^*\}$ is added to $\mathcal{B}$. Then $\ell_{t+1}=b$.

Suppose there does exist a $Y\in\mathcal{B}$ at the beginning of iteration $t$ such that
$\texttt{bin}(Y)=b$. $Y$ is not cost-effective by definition of $\ell_t$.
Then if $b=r$,
\begin{align*}
  c(X\cup\{x^*\}) \leq (\ln(1/\epsilon) + 1)c(A^*) < c(Y)
\end{align*}
and so $Y\prec X\cup\{x^*\}$. Then $Y$ is replaced with $X\cup\{x^*\}$ in $\mathcal{B}$.
If $b<r$, and recalling that $b\neq 0$ as explained at the beginning of the proof, then
\begin{align*}
\phi(X\cup\{x^*\}) \leq c(A^*) < \phi(Y).
\end{align*}
Therefore again $Y\prec X\cup\{x^*\}$, so $Y$ is replaced with $X\cup\{x^*\}$ in $\mathcal{B}$.
In both cases, $\ell_{t+1}=b$.

\paragraph{Part Five.}
Finally, the expected number of iterations until $\ell_t=r$ is analyzed. Once $\ell_t=r$,
the solution in $\mathcal{B}$ mapping to bin $r$ satisfies the
conditions of the theorem statement.

Suppose it is the beginning of iteration $t$ of EASC such that
$\ell_t < r$.
Then with probability at least
\begin{align*}
  \frac{1}{|\mathcal{B}|}\frac{1}{n}\left(1-\frac{1}{n}\right)^{n-1}
  \geq \frac{1}{en(r+1)}
\end{align*}
the set $X$ corresponding to $\ell_t$ will be chosen and mutated into $X\cup\{x^*\}$.
By Part Four, if this occurs $\ell_t < \ell_{t+1}$.
By Part One, the sequence is non-decreasing.
This means that the expected number of steps for $\ell_t$ to reach $r$ is at most
$enr(r+1)$. The bound on $r$ given by Proposition \ref{proposition:bins} of Section \ref{section:bin}
gives the theorem statement.
\end{proof}




\section{Experimental Analysis}
\label{section:experiments}
\begin{figure*}
  \centering
  \hspace{-10pt}
  \subfigure[][ca-GrQc $\tau=250$] {
    \label{fig:grqc}
    \includegraphics[width=0.25\textwidth]{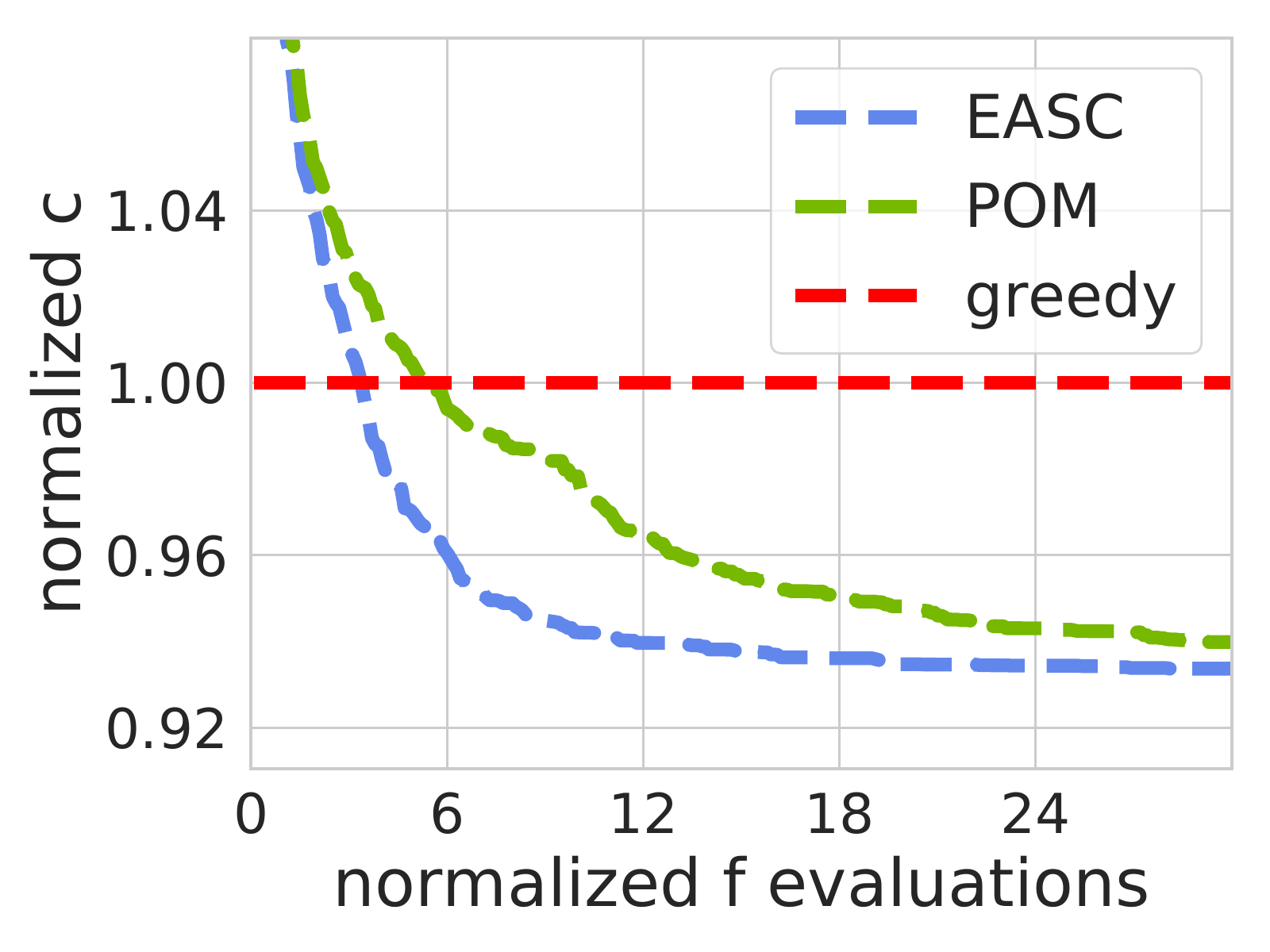}
  }
  \hspace{-10pt}
  \subfigure[][ca-HepPh $\tau=970$] {
    \label{fig:hepph}
    \includegraphics[width=0.25\textwidth]{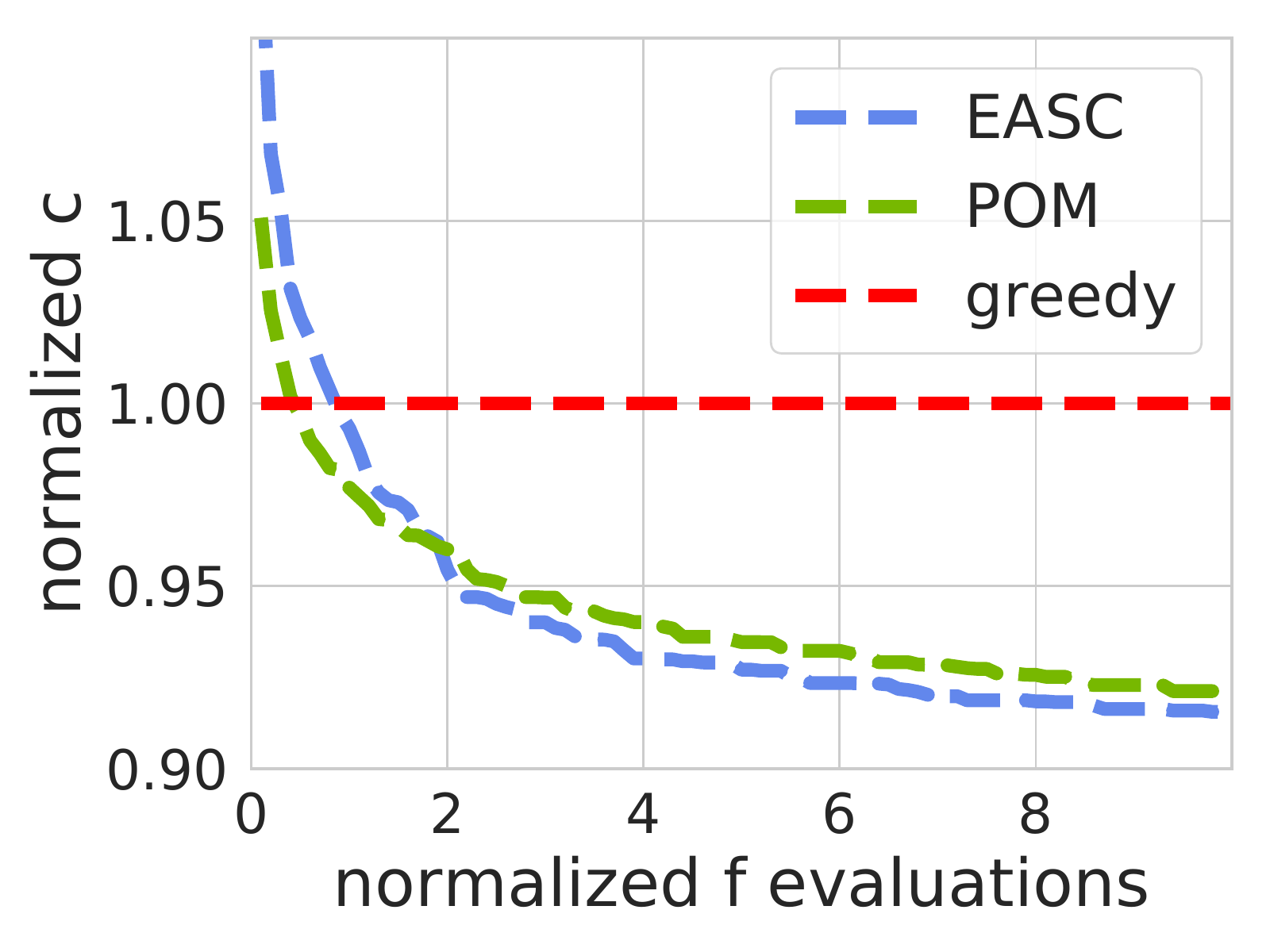}
  }
  \hspace{-10pt}
  \subfigure[][wiki-Vote $\tau=650$] {
    \label{fig:vote}
    \includegraphics[width=0.25\textwidth]{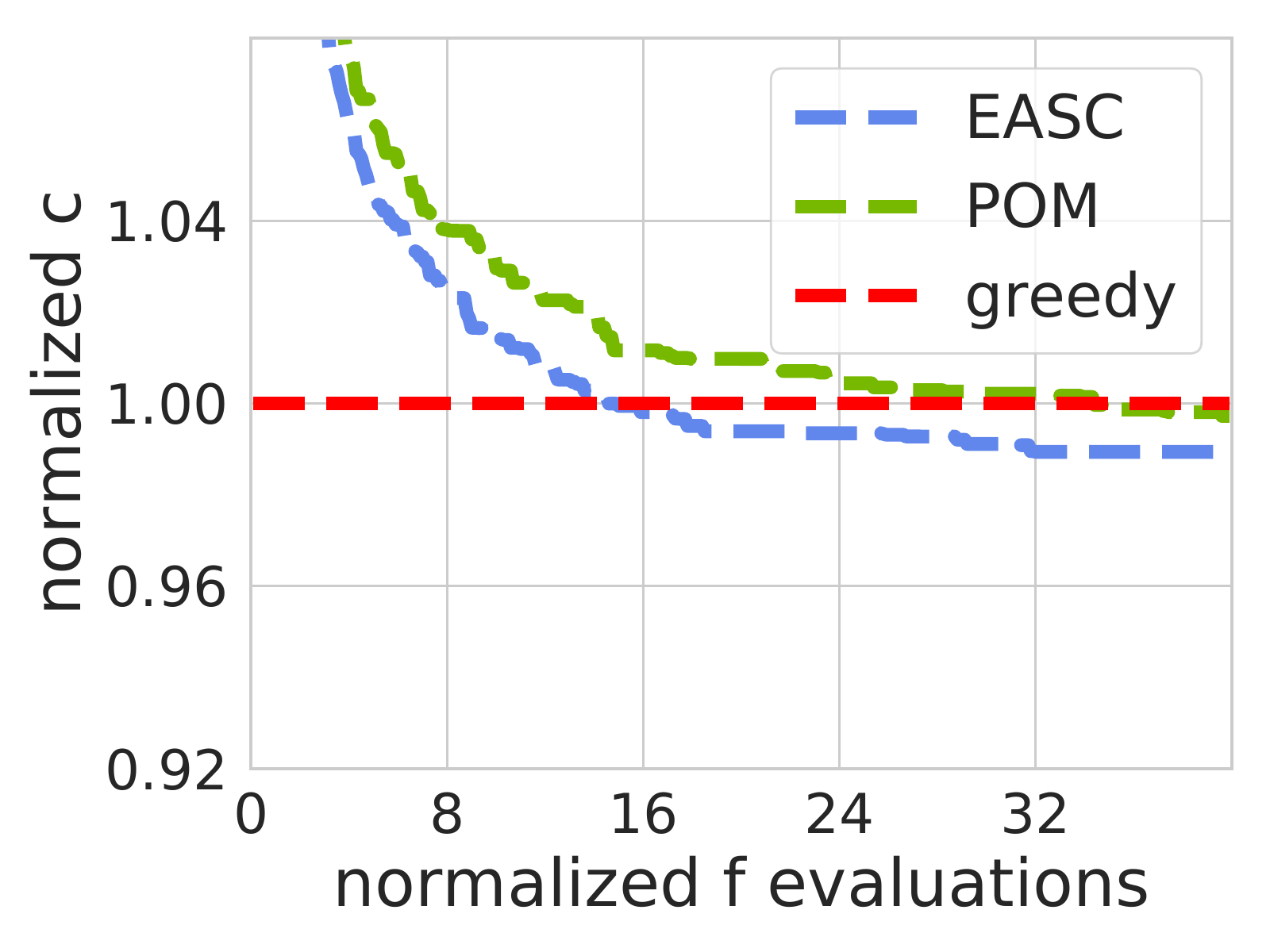}
  }
  \hspace{-10pt}
  \subfigure[][ego-Facebook $\tau=410$] {
    \label{fig:facebook}
    \includegraphics[width=0.25\textwidth]{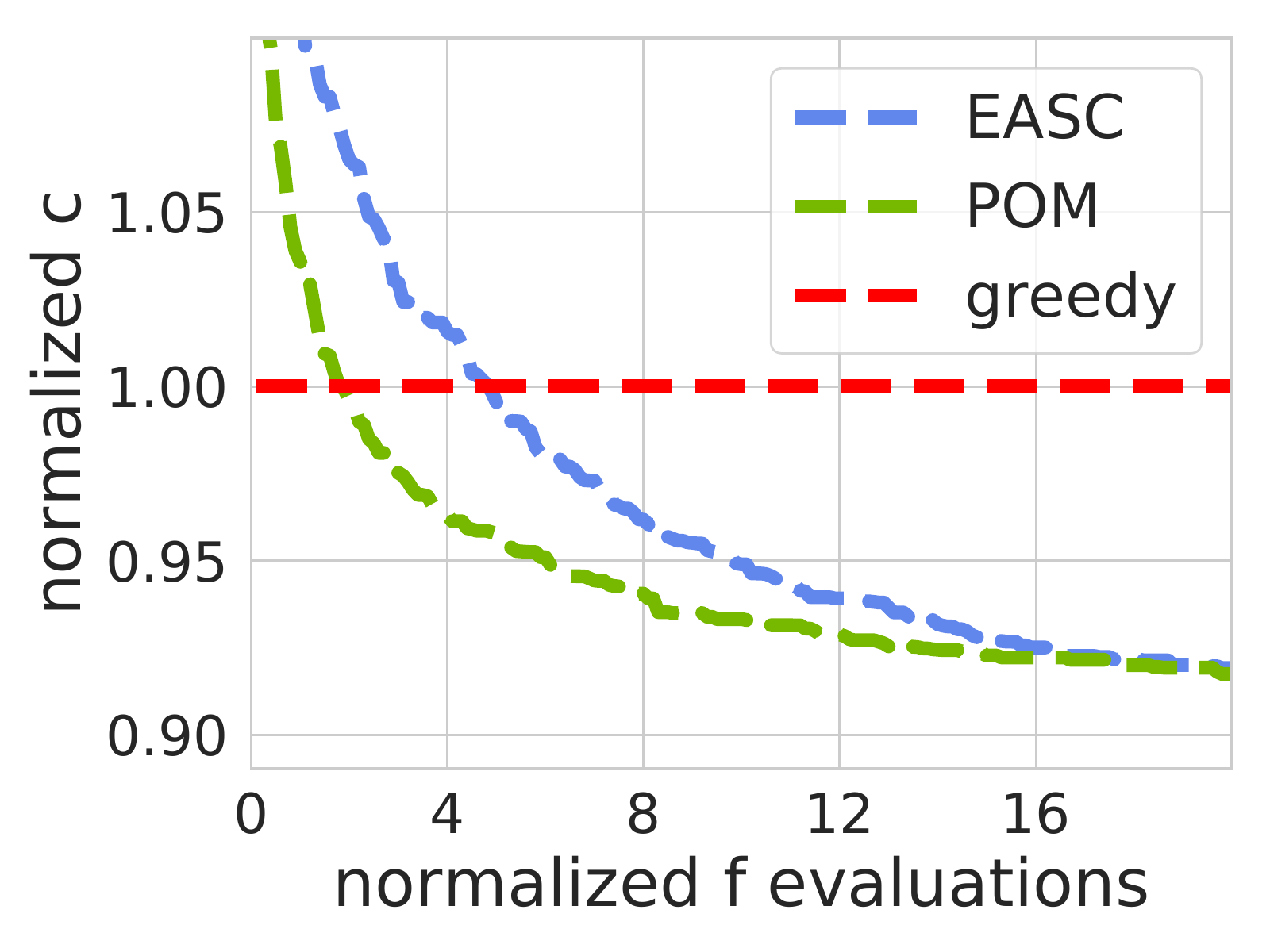}
  }
  \caption{
Over the duration of EASC and POM,
the minimum cost of any solution in the population with $f$
value above $(1-\epsilon)\tau$ is plotted. Both cost and number of $f$ evaluations
are normalized by that of the greedy algorithm.
  }
  \label{entirefig}
\end{figure*}
In this section, EASC is experimentally evaluated on instances of the
Influence Threshold Problem (IT)
\cite{goyal2013}, a special case of MCSC. EASC is compared to the greedy
algorithm and POM \cite{qian2015constrained}.
In all experiments, EASC and POM find solutions of lower cost than the greedy algorithm,
and in most cases EASC converges faster than POM to a low cost.
Code to run the experiments is publicly available at \url{https://gitlab.com/vcrawford/easc.git}.

\subsection{Application and Setup}
The experiments are run on instances of the Influence Threshold Problem,
defined as follows.
Let $G=(V,E)$ be a social network where vertices $V$ represents users, and
directed edges $E$ represent social connections. Activation of users in
the social network starts from an initial seed set and then propagates
across ``live edges''
according to the independent cascade model \cite{kempe2003maximizing},
in which every edge $(u,v)\in E$ has an independent probability $p_{uv}$ of being live.

For every user $v\in V$, there is a cost $c_v$ of seeding that user.
The cost of seeding a set $X$ is $c(X)=\sum_{x\in X}c_x$.
The function $f(X)$
is the expected number of users that will become active if $X$ is
seeded; $f$ is monotone submodular \cite{kempe2003maximizing}.
Then the IT problem is defined as follows: given an activation threshold $\tau$, find
$\text{argmin}\{c(X): X\subseteq S, f(X) \geq\tau\}$.


The experiments are run on four real social networks from SNAP \cite{leskovec2015snap}:
ca-GrQc ($n=5242$),
ca-HepPh ($n=12008$),
wiki-Vote ($n=7115$), and
ego-Facebook ($n=4039$).
The independent cascade model is used to model activation from a seed set
for the above four social networks with constant edge probabilities
$p=0.07$, $p=0.02$, $p=0.04$, and $p=0.013$, respectively.

Computing the expected activation $f( X )$ under the independent cascade
model is \#P-hard \cite{chen2010}. Instead of evaluating $f$ directly,
the reverse influence sampling approach \cite{borgs2014maximizing} with 100,000 samples
is used in order to approximate $f$.
The same set of samples is used for all algorithms on a data set.

The cost function $c(X)=\sum_{x\in X}c_x$,
where every node $v$ in the social network is assigned a cost $c_v$ that is its outgoing degree perturbed by random
multiplicative noise\footnote{Under this model of cost,
social network users with more outgoing edges are generally more expensive to seed,
but individual preferences factor into the price of seeding via random noise.}
In particular, node $v$ with outdoing degree $d$ has cost $1+(1+|\xi|)d$
where $\xi\sim \mathcal{N}(\mu=0,\sigma=0.5)$\footnote{A normal distribution
with mean $0$ and standard deviation $0.5$.}.

The instance of MCSC for each social network is run with a different threshold $\tau$.
The greedy algorithm is run with input $\tau$ and $\epsilon=0.05$.
EASC is run with $\tau$, $\epsilon=0.05$, and $\delta=1-c_{min}/B$ where
$B$ is the cost of the output of the greedy algorithm when run with
$\tau$ and $\epsilon=0$. 
POM is run with threshold $(1-\epsilon)\tau$ for
fair comparison with EASC, although POM is not
a bicriteria algorithm.
EASC and POM are run 3 times on each instance, and the results are averaged.

\subsection{Results}
The experimental results are plotted in Figure \ref{entirefig}.
At small intervals over the duration of EASC and POM,
the minimum cost of any solution in the population with $f$
value above $(1-\epsilon)\tau$ is plotted. The cost and the number of $f$
evaluations are normalized by that of the greedy algorithm. That is, if the greedy
algorithm returned a set $G$, then the costs plotted are normalized by
$c(G)$ and the number of function evaluations are normalized by
$n|G|$.
%

In all experiments, EASC and POM were able to find a better solution than the greedy algorithm.
In ca-HepPh (Figure \ref{fig:hepph}), both EASC and POM find a better solution than the greedy
algorithm \textit{in less $f$ evaluations}. These results demonstrate
an ability to
improve on the solution quality of the greedy algorithm.

In ca-GrQc (Figure \ref{fig:grqc}) and wiki-Vote (Figure \ref{fig:vote}), EASC finds a better
solution more quickly than POM throughout the entire experiment. In contrast, in
ca-HepPh (Figure \ref{fig:hepph}) and ego-Facebook (Figure \ref{fig:facebook}) POM finds a better
solution more quickly in the earlier stages of the experiment, but EASC catches up and either
outperforms POM for the remainder of the iterations (caHepPh) or converges to about the same performance.
This behavior may be explained by the fact that at first POM has a smaller population since every
solution competes with every other solution, unlike EASC where competition is restricted to within
bins. But as the population of POM gets larger (at least $3$
times larger than EASC at the end of these
experiments), the improvement of POM is slowed.

%


\section*{Acknowledgements}
Victoria G. Crawford was supported by a Harris Corporation Fellowship.
Alan Kuhnle provided helpful feedback in preparation of the manuscript.
\clearpage
\bibliography{arxiv}
\bibliographystyle{abbrvnat}
\clearpage

  \section{Appendix}
  	Background and results that are not included in the paper due to space constraints are provided here.
  	In particular, the greedy algorithm and its approximation guarantee for MCSC is
  	discussed in Section \ref{appendix:greedy}.
  	Lemmas needed for the proof of Theorem \ref{theorem:main} are proven in
  	Section \ref{appendix:lemmas}.

    \subsection{The Greedy Algorithm}
    \label{appendix:greedy}
    Pseudocode for the greedy algorithm is provided in
    Algorithm \ref{algorithm:greedy}. It should be noted that the classic
    greedy algorithm is Algorithm \ref{algorithm:greedy} with $\epsilon=0$,
    but this alternate version is considered since it is more comparable
    to EASC.

    As Proposition \ref{proposition:greedy} states,
    the bicriteria approximation ratio proven for EASC in Section
    \ref{section:approximation} also holds for the greedy algorithm.
    This result is not novel to this paper, but was proven in the context of
    influence by \citeauthor{goyal2013} (\citeyear{goyal2013}).
    The proof is easily adjusted to hold for general MCSC, and the proof is
    included here.

    \begin{algorithm}[tb]
      \caption{Greedy Algorithm}
      \label{algorithm:greedy}
      \begin{algorithmic}
        \STATE {\bfseries Input:} MCSC instance parameters $f:2^S\to\mathbb{R}_{\geq 0}$, $c:2^S\to\mathbb{R}_{\geq 0}$, and $\tau$, and $\epsilon\in (0,1)$.

        $A=\emptyset$

        \WHILE {$f(A) < (1-\epsilon)\tau $}
        \STATE {
          $u = \text{argmax}_{x\in S}\Delta f_{\tau}(A, x)/c(x)$

          $A = A \cup \{u\}$
        }
        \ENDWHILE

        \RETURN $A$

        \end{algorithmic}
      \end{algorithm}

      \begin{proposition}
        \label{proposition:greedy}
        Suppose that we have an instance of MCSC with optimal solution $A^*\neq\emptyset$, and run
        Algorithm \ref{algorithm:greedy} with
        input $\epsilon > 0$.
        Then the set $A$ returned satisfies $f(A)\geq(1-\epsilon)\tau$
        and
        \begin{align*}
          c(A) \leq \left(\ln\left(\frac{1}{\epsilon}\right)+1\right)c(A^*).
        \end{align*}
      \end{proposition}
      \begin{proof}
        The feasibility guarantee is clear from the stopping condition on Algorithm \ref{algorithm:greedy}.
        Let $A=\{a_1,...,a_k\}$ in the order of being chosen. Denote by $A_i=\{a_1,...,a_i\}$.
        By Lemma \ref{lemma:gain}, for any $i\in\{1,...,k\}$
        \begin{align*}
          \tau - f_{\tau}(A_{i}) &\leq \left(1-\frac{c(a_{i})}{c(A^*)}\right)(\tau-f_{\tau}(A_{i-1})) \\
          &\leq \exp\left(\frac{c(a_i)}{c(A^*)}\right)(\tau-f_{\tau}(A_{i-1})).
        \end{align*}
        By induction it is then the case that
        \begin{align}
          \tau - f_{\tau}(A_{k-1}) \leq \exp\left(\frac{c(A_{k-1})}{c(A^*)}\right)\tau.
          \label{eqn:31brewhj}
        \end{align}
        Algorithm \ref{algorithm:greedy} did not return $A_{k-1}$, and so
        $\tau - f_{\tau}(A_{k-1}) > \epsilon\tau$. Applying this to Equation
        \ref{eqn:31brewhj} and then re-arranging gives
        \begin{align*}
          c(A_{k-1})\leq \ln\left(\frac{1}{\epsilon}\right)c(A^*).
        \end{align*}
        By applying Lemma \ref{lemma:lastelt}, it is the case that
        \begin{align*}
          c(A) = c(A_{k-1}) + c(a_k) \leq \left(\ln\left(\frac{1}{\epsilon}\right)+1\right)c(A^*).
        \end{align*}
      \end{proof}

    \subsection{Lemmas}
    \label{appendix:lemmas}
      Lemmas needed for the proof of Theorem \ref{theorem:main} of Section
      \ref{section:approximation} are proven in this section.
      Lemma \ref{lemma:gain} is a slight variation of a commonly used result
      and is not novel to this paper
      \cite{nemhauser1978analysis}.
      Lemma \ref{lemma:lastelt} is assumed without proof in the proof of Proposition
      \ref{proposition:greedy} by \citeauthor{goyal2013}, but is proven here
      for clarity.

    \paragraph{Lemma \ref{lemma:gain}}
      Suppose that we have an instance of MCSC with optimal solution $A^*\neq\emptyset$.
      Let $X\subseteq S$ and $x^* = \text{argmax}_{x\in S}\Delta f_{\tau}(X,x)/c(x)$.
      Then
      \begin{align*}
        \tau - f_{\tau}(X\cup\{x^*\}) \leq \left(1-\frac{c(x^*)}{c(A^*)}\right)(\tau-f_{\tau}(X)).
      \end{align*}
    \begin{proof}
      \setcounter{equation}{0}
      Define an arbitrary order on the elements of $A^*$, $y_1,...,y_m$.
      It is the case that
      \begin{align*}
        c(A^*)\frac{\Delta f_{\tau}(X,x^*)}{c(x^*)}
        = \sum_{i=1}^{m}c(y_i)\frac{\Delta f_{\tau}(X,x^*)}{c(x^*)}
        \geq \sum_{i=1}^{m}\Delta f_{\tau}(X,y_i)
      \end{align*}
      by definition of $x^*$. Now, if $f$ is monotone submodular, then
      $f_{\tau}$ is as well, which implies
      \begin{align*}
        \sum_{i=1}^{m}\Delta f_{\tau}(X,y_i)
        &\geq \sum_{i=1}^{m}\Delta f_{\tau}(X\cup\{y_1,...,y_{i-1}\},y_i)\\
        &= \tau - f_{\tau}(X).
      \end{align*}
      Therefore,
      \begin{align*}
        c(A^*)\frac{\Delta f_{\tau}(X,x^*)}{c(x^*)} \geq \tau - f_{\tau}(X),
      \end{align*}
      which can be re-arranged to get the statement of the lemma.
    \end{proof}

    \paragraph{Lemma \ref{lemma:lastelt}}
      Suppose that we have an instance of MCSC with optimal solution $A^*\neq\emptyset$.
      Let $X\subseteq S$ such that $f(X) < \tau$
      and $x^* = \text{argmax}_{x\in S}\Delta f_{\tau}(X,x)/c(x)$.
      Then $c(x^*)\leq c(A^*)$.
    \begin{proof}
      First, it will be shown that there exists an $a^*\in A^*$ such that
      \begin{align*}
        \frac{\Delta f_{\tau}(X,a^*)}{c(a^*)} \geq \frac{\Delta f_{\tau}(X,A^*)}{c(A^*)}.
      \end{align*}
      Suppose no such $a^*\in A^*$ existed.
      Then
      \begin{align*}
        \sum_{a\in A^*}\Delta f_{\tau}(X,a)
        < \sum_{a\in A^*}\frac{c(a)}{c(A^*)}\Delta f_{\tau} (X, A^*)
        = \Delta f_{\tau} (X, A^*)
      \end{align*}
      which implies that $f_{\tau}$ is not monotone submodular. However,
      this is a contradiction because $f$ being monotone submodular implies that
      $f_{\tau}$ is monotone submodular. Therefore such an $a^*$ must exist.
%

      $f(X)<\tau$ implies that $\Delta f_{\tau}(X,A^*) > 0$, which in turn
      implies that $\Delta f_{\tau}(X,a^*) > 0$. In addition, $f(X)<\tau$ and the
      submodularity of $f$ implies that $\Delta f_{\tau}(X,x^*) > 0$.
      Then
      \begin{align*}
        c(x^*) \leq \frac{\Delta f_{\tau}(X,x^*)}{\Delta f_{\tau}(X,a^*)}c(a^*)
        \leq \frac{\Delta f_{\tau}(X,x^*)}{\Delta f_{\tau}(X,A^*)}c(A^*)
        \leq c(A^*)
      \end{align*}
      since $\Delta f_{\tau}(X,x^*) \leq \Delta f_{\tau}(X,A^*)$.
    \end{proof}

\end{document}